\documentclass[12pt]{iopart}
\usepackage{graphicx}
\usepackage{capt-of}
\usepackage{amssymb}
\usepackage{lineno}

\def \FIGCOLORTYPE {_col}

\newcommand{\ie}{{\textit i.e.}}

\begin{document}

\title{APOLLO clock performance and normal point corrections}

\author{Y.~Liang$^1$, T.W.~Murphy,~Jr.$^1$, N.R.~Colmenares$^1$, J.B.R.~Battat$^2$}
\address{$^1$ Center for Astrophysics and Space Sciences, University of California, San Diego, 9500 Gilman Drive, La Jolla, CA 92093-0424, USA}
\address{$^2$ Department of Physics, Wellesley College, 106 Central St, Wellesley, MA 02481, USA}
\eads{\mailto{tmurphy@physics.ucsd.edu}}

\maketitle

\begin{abstract}

The Apache Point Observatory Lunar Laser-ranging Operation (APOLLO) has
produced a large volume of high-quality lunar laser ranging (LLR) data
since it began operating in 2006.  For most of this period, APOLLO has
relied on a GPS-disciplined, high-stability quartz oscillator as its
frequency and time standard.  The recent addition of a cesium clock as part
of a timing calibration system initiated a comparison campaign between the
two clocks.  This has allowed correction of APOLLO range
measurements---called normal points---during the overlap period, but also
revealed a mechanism to correct for systematic range offsets due to clock
errors in historical APOLLO data.  Drift of the GPS clock on $\sim 1000$~s
timescales contributed typically 2.5~mm of range error to APOLLO
measurements, and we find that this may be reduced to $\sim 1.6$~mm on
average.  We present here a characterization of APOLLO clock errors, the
method by which we correct historical data, and the resulting statistics.

\end{abstract}

\section{Introduction \label{sec:intro}}

Lunar laser ranging (LLR) has provided many of the best tests of
fundamental gravity since reflectors were first placed on the Moon in 1969
\cite{Bender:1973zz,Dickey:1994zz,murphyLLRReview2013}.  Measurements are
packaged into ``normal points," primarily composed of a representative epoch
and round-trip travel time for a contiguous set of single-shot measures to
a single reflector spanning several minutes of time.  Initial range
precision from the McDonald 2.7~m Telescope hovered around 200~mm---based
on a few-nanosecond pulse width using a ruby laser \cite{Bender:1973zz}.
In the mid 1980's, new efforts in France, Texas, and Hawaii took advantage
of technology improvements to produce 20--30~mm range precision even though
utilizing 1-meter-class apertures \cite{Dickey:1994zz}.  In 2006, the
Apache Point Observatory Lunar Laser-ranging Operation (APOLLO) on a 3.5~m
telescope began producing LLR data with an estimated precision of a few
millimeters \cite{murphyAPOLLO2008,battatPASP2009}.  Given that
relativistic effects in the lunar orbit, as evaluated in the solar system
barycenter frame, have $\sim$10\,m amplitude, millimeter measurements can
constrain relativistic gravity at the 0.01\% level.

The time and frequency standard originally chosen for APOLLO was a TrueTime
XL-DC model employing a high stability ovenized quartz oscillator
disciplined by reference to the global positioning system (GPS).  Low phase
noise ($-153$~dBc at frequencies $> 1$~kHz) translates to a time-domain jitter of
4.0~ps over 2.5~s intervals---the interval associated with lunar round-trip
travel time.  This corresponds to a frequency offset of $1.6\times
10^{-12}$.  After multiplying the native 10~MHz frequency by five to create
a 50~MHz system clock for APOLLO, the $20\log 5 = 14$~dB hit to phase noise
translates to 7~ps of jitter, or approximately one millimeter of one-way
distance.  Averaging over minutes can, in principle, reduce this to well
below one millimeter.

The XL-DC steering algorithm applies a Kalman filter over the previous
$\sim 2000$~s to assess the frequency offset between the GPS solution and
the local quartz oscillator, then computes a correction voltage to steer
the oscillator via a digital-to-analog converter (DAC).  A new DAC value is
computed every 10~s.  The APOLLO system monitoring software queries and
stores statistics provided by the clock on a 10~s cadence.  These data
include phase offset (typically $\sim 10$~ns), frequency offset (typically
$<10^{-11}$), drift rate (typically $-10^{-11}$ per day), and DAC integer value
(migrating from $-1400$ in 2006 to $-4900$ in 2017), and are essentially
continuously recorded through the entire history of the APOLLO experiment.

The DAC necessarily produces discrete steps, and the size of the steps is
designed to balance lifetime oscillator drift, intrinsic noise/errors, and
bit-depth of the DAC.  For the APOLLO XL-DC, one DAC step corresponds to
a fractional frequency shift of $1.2\times 10^{-11}$.  The corresponding
2.5~s LLR measurement therefore jumps by 30~ps, or 4.5~mm at each DAC step.
This is larger than is ideal for a millimeter-scale operation, but the
root-mean-square (RMS) effect on LLR measurements is reduced by a factor of
$\sqrt{12}$ to 1.3~mm if the ``truth'' frequency is uniformly distributed
between adjacent DAC values.  Observation of the trend in the recorded DAC
value---spending many minutes or hours at the same value or sometimes
dithering between adjacent steps---appeared to justify this assumption.  As
such, we have, until now, accounted for clock error by adding 10~ps in
quadrature with the statistical uncertainty of APOLLO normal points.

In February 2016, APOLLO installed a cesium clock (Microsemi 5071A) as a
first phase of a new absolute calibration system
\cite{ACS2017}, which became fully operational in September 2016.  A
Universal Counter (UC; Agilent 53132A) began collecting essentially
continuous measurements of the XL-DC frequency with reference to the Cs
10~MHz output on 10~s intervals to $10^{-13}$ resolution.  As such, we are
able to measure and correct for XL-DC clock errors. Further, we now have a
basis against which to judge the nature and quality of the XL-DC's
self-reported statistics.  Because we found that the recorded statistics
bear resemblance to the ``truth'' data from the Cs clock, we are able to
back-correct archival APOLLO data based on the historical log of
XL-DC-reported clock statistics, thereby reducing the larger-than-expected
influence of the GPS-disciplined clock.

In Section~\ref{sec:clock_comparison}, we summarize the characteristics of the two
clocks.  In Section~\ref{sec:np-correct}, we use the UC comparison data to correct APOLLO data in 2016
when the GPS clock was still used as APOLLO's frequency standard, but
direct comparison to the Cs clock was available.  Finally, in Section~\ref{sec:correction}, we present an
algorithm to predict/approximate the UC ``truth'' data using the GPS clock
statistics and characterize the impact of these corrections on APOLLO
normal points.

\section{Clock Comparison \label{sec:clock_comparison}}

\begin{figure}[t]
  \centering
  \includegraphics[width=0.7\textwidth]{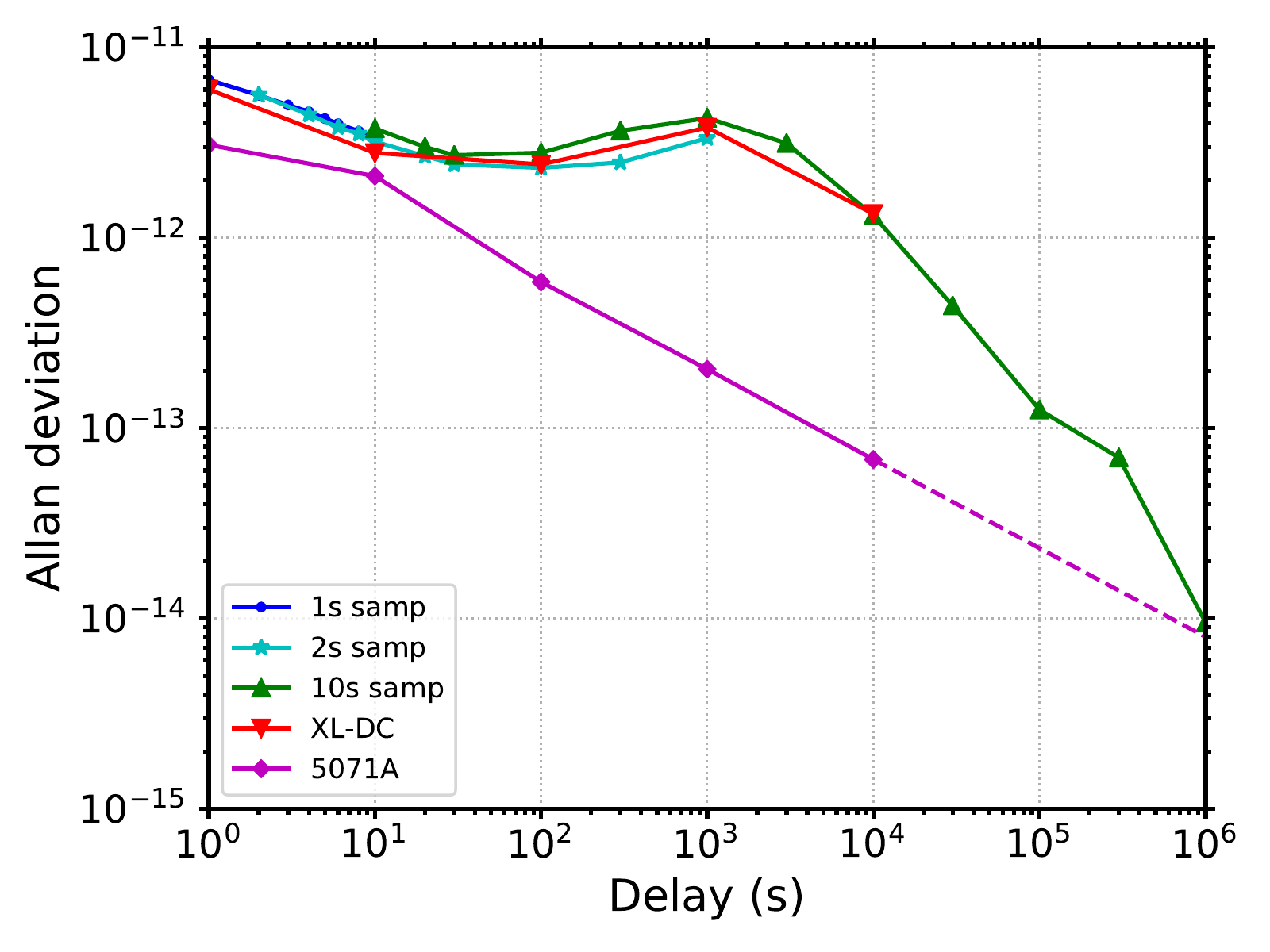}
  \caption{\label{fig:allan} Clock Allan deviation measurements.  Cs clock
statistics (magenta; lower), provided by the manufacturer, were measured against a
hydrogen maser.  The dashed line projects the measurements to longer times.
The XL-DC Allan deviation was measured with the UC at 1\,s, 2\,s, and 10\,s
averaging (blue, cyan, green, respectively), using the Cs clock as a
reference.  To estimate the Allan Deviation of the XL-DC clock alone, the
red points show the quadrature subtraction of the Cs clock deviation
(magenta) from the average sampled data.  Assuming that the Cs
clock projection is accurate, the Cs clock is more stable than the XL-DC
for time intervals less than 2 weeks. Beyond that, the GPS-based
disciplining of the XL-DC improves its stability relative to the
free-running Cs standard. The degradation of the XL-DC stability at
$10^3$\,s is likely due to atmospheric fluctuations that burden the GPS
solution.}.
\end{figure}

Compared to the GPS-steered (XL-DC) clock, the Cs clock provides superior
frequency stability on intermediate timescales ($\sim 10^3$~s) and can be
used to characterize the GPS clock. The Allan deviations for the two clocks
can be seen in Figure~\ref{fig:allan}.  Precision at the shortest intervals
($\leq 10$\,s) is fairly similar---both using quartz oscillators.  At long
times ($> 10^6$\,s), the GPS clock is better due to the tie, via GPS, to an
ensemble of atomic clocks held to the international definition of time at
sea level.  At intermediate times, around $10^3$~s, the GPS clock is
afflicted by atmospheric variation (largely water vapor) impacting the GPS
solution.  We also occasionally (on a roughly weekly basis) interrupt the
frequency measurement campaign for about one minute to check the phase
drift between the two clocks' pulse-per-second outputs.  The net effect of
the built-in Cs offset ($-1.3\times 10^{-13}$, as measured by the
manufacturer against a hydrogen maser) and gravitational redshift at the
site ($+3.0\times 10^{-13}$) combine to make the Cs clock run faster than
the sea-level-tied XL-DC by $\sim 15$~ns per day.

\section{Clock Corrections to Normal Points \label{sec:np-correct}}

The Cs clock became operational at the observatory in mid-February 2016,
and became the primary APOLLO frequency reference at the beginning of 2017.
In the intervening time, the XL-DC was still the frequency standard for APOLLO's
lunar range measurements, while the UC collected measurements of its frequency relative
to the Cs clock.  This permits a correction to LLR normal points
based on the frequency offsets measured during range acquisition.

\begin{figure}[t]
  \centering
  \includegraphics[width=0.7\textwidth]{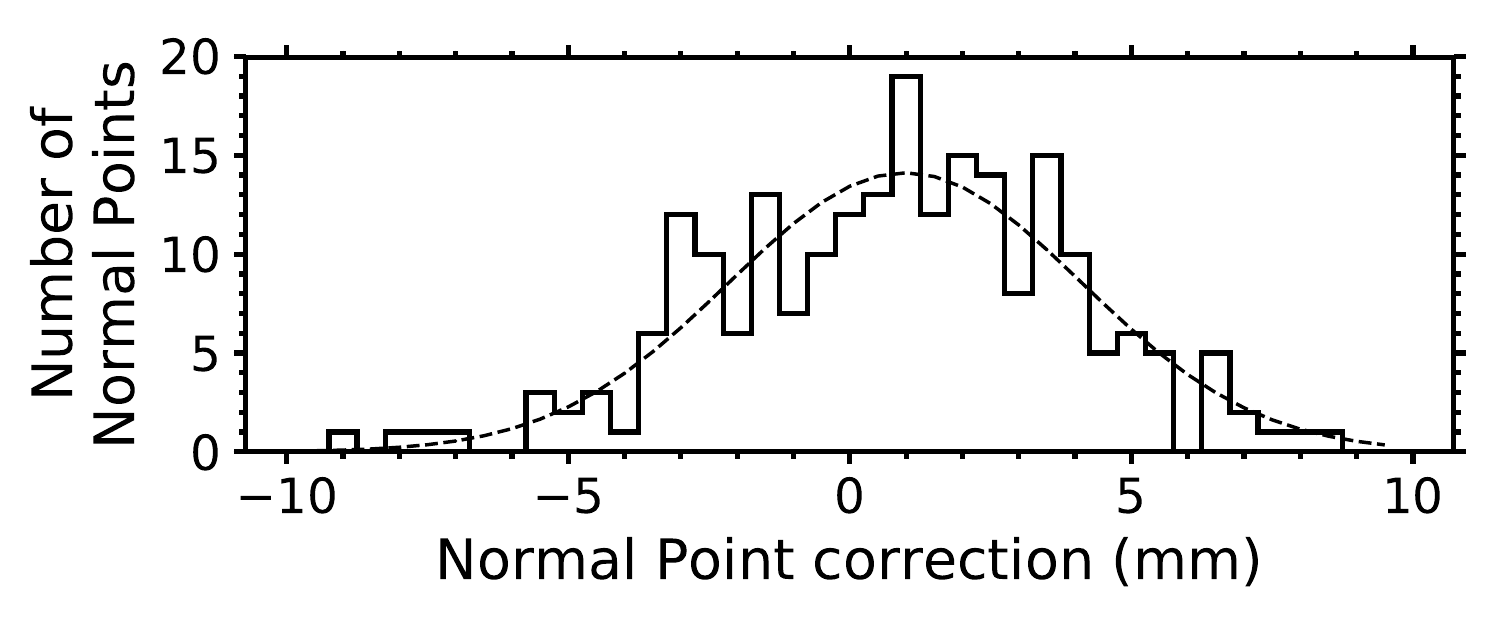}
  \caption{\label{fig:clk-cmp} Corrections to APOLLO normal points
acquired between 2016 Feb. 16 and 2016 December 12, in 0.5\,mm bins.  The corrections are based on the UC
measurements of the GPS-disciplined clock offsets relative to the Cs clock.  We find an offset of 0.9\,mm and standard deviation of 3.2\,mm
(3.1\,mm according to Gaussian fit).}
\end{figure}

Normal points acquired between 2016 February 16 and the end of 2016 are
therefore correctable in this manner. The statistics of the corrections are
seen in Figure~\ref{fig:clk-cmp}, indicating typical clock errors on the
scale of 3\,mm.  The clock comparison measurement indicates that the
clock-induced error is twice as large as had been appreciated, as motivated
in Section~\ref{sec:intro}.  APOLLO normal points of the past should have
had a $\sim 20$~ps (3~mm) uncertainty term applied in quadrature, rather
than the 10~ps that was habitually applied.  However, because we now have
methods to measure and correct the clock errors, we will not need to modify
uncertainties as drastically.

The UC data are averaged over 10~s intervals, and according to
Figure~\ref{fig:allan}, we might expect the GPS clock to produce a standard
deviation approximately equal to the Allan deviation at 10~s, or in this
case approximately $4\times 10^{-12}$.  Indeed this is close to the
observed intrinsic scatter in the UC data (more on this in
Section~\ref{sec:back-correct}).  Averaged over typical normal point
durations of 150--250~s, and thus 15--25 clock comparison measurements, we
might expect a resultant determination of the clock performance for a
normal point to be good at the $10^{-12}$ level, translating to $\sim
0.4$~mm in one-way range.  Thus, the corrections described in this section
may be considered to be adequate at the sub-millimeter level.

\section{Back-correcting historical APOLLO data
\label{sec:correction}}

The frequency difference between the 10\,MHz signals from the XL-DC clock
and the Cs clock are measured to 1\,$\mu$Hz resolution ($10^{-13}$) every
10~s. Taking for now the UC measurements of the XL-DC frequency to
represent ``truth,'' an algorithm is developed using the XL-DC's
self-reported frequency offset and DAC values to ``predict'' the frequency
offsets as measured by the UC.  Then the algorithm is applied to historical
APOLLO data and associated statistics are presented.

\subsection{Back-correction algorithm \label{sec:back-correct}}

\begin{figure}[t]
  \centering
  \includegraphics[width=0.7\textwidth]{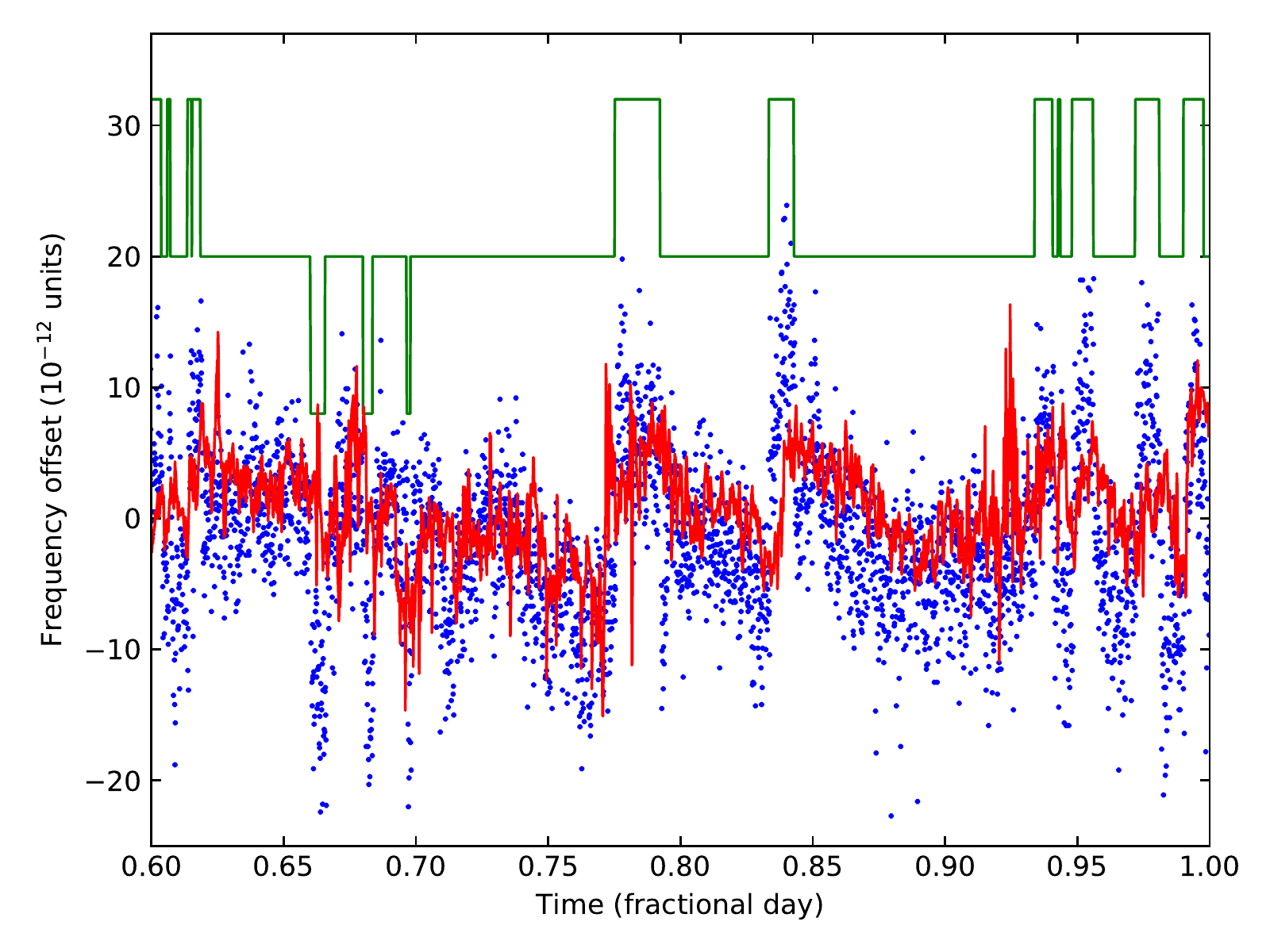}
  \caption{\label{fig:uc-original} Comparison of XL-DC reported frequency
offset (red jagged line) and DAC (green at top, offset arbitrarily) to the
UC measurements of XL-DC frequency (blue dots). The goal of the algorithm is to use the offset and the DAC information to best approximate the UC data. }
\end{figure}

\begin{figure}[t]
  \centering
  \includegraphics[width=0.7\textwidth]{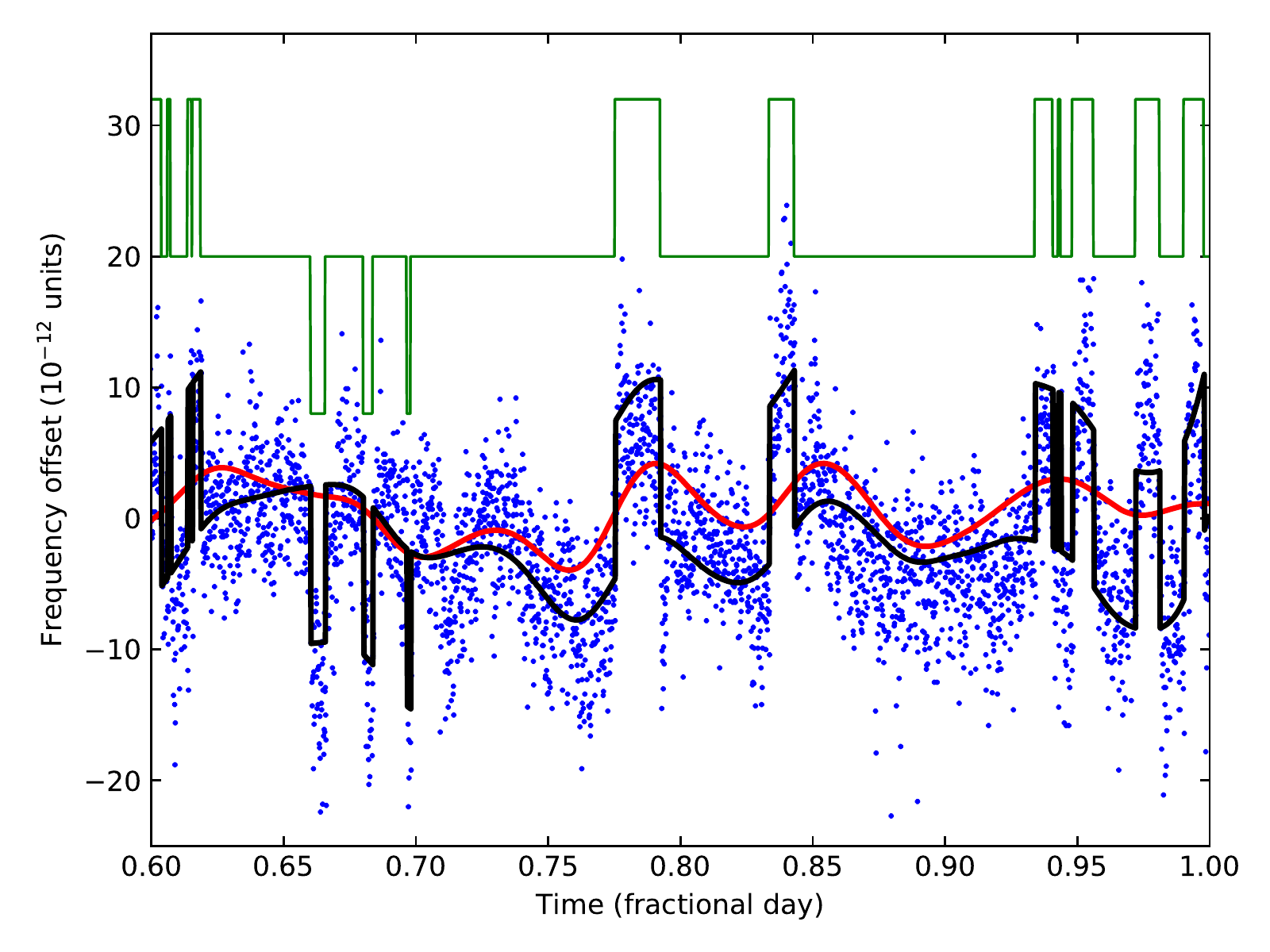}
  \caption{\label{fig:uc-prediction} An example prediction result on a test
data set (not part of the training set; same period as in
Figure~\ref{fig:uc-original}). The blue dots are UC measurements; the red thin line is the
low-pass-filtered frequency offset; and the black thick line is the
prediction result. In terms of error reduction, the mean squared error in
the example is reduced from 71.5 (total UC variance) to 45.2 (post-fit
variance), in units of $10^{-24}$. }
\end{figure}

Figure~\ref{fig:uc-original} shows $\sim$10\,hours of measurements of the
frequency difference between the Cs and GPS clocks, as well as the
XL-DC-reported frequency offset and DAC value.  The DAC steps are evident
in the UC data, and excursions in frequency appear in similar form in the
UC data. This suggests an algorithmic approach to reconstructing a
facsimile to the UC data based only on the XL-DC-reported data.
Specifically, DAC steps can be preserved by applying a high-pass filter
with $\sim 1$\,h time constant to the DAC data. Conversely, a low-pass
filter applied to the frequency offset ($\sim 0.25$~h time constant),
followed by deconvolution and shifting can reasonably match
non-DAC-generated features in the UC data. These separable influences are
added to form a prediction. The relevant parameters such as the scaling
factors and the filter frequencies are optimized through a least-squares
fit to the UC data. After training the algorithm on 235 days of joint UC
and XL-DC data, a parameter set that minimizes the mean squared difference
between the UC data and the prediction was determined.
Figure~\ref{fig:uc-prediction} provides an example result.

In order to properly interpret the goodness of fit, we need to characterize
the intrinsic UC variations absent influences from GPS steering, since the
intrinsic UC noise contributes to both the total UC variance and to the
post-fit variance (the difference between UC and the prediction). More
precisely, the UC intrinsic variance should be subtracted from both the
pre-fit and post-fit variance to better elucidate the real structures. The
amplitude of intrinsic noise in the UC measurements could be determined by
investigation of “silent” periods, where the UC, offset and DAC are flat.
Statistics based on 85 silent periods revealed a tight distribution of UC
variance centered at 16 (measured in $10^{-24}$ units), corresponding to a
standard deviation of $4\times 10^{-12}$---consistent with the Allan
deviation shown in Figure~\ref{fig:allan}.  Thus, to convert a variance of
pre-fit or post-fit data into a corresponding range contribution, we first
subtract the quiescent variance, take the square root, and multiply by
lunar distance (or round-trip time times half the speed of light).  For the
example in Figure~\ref{fig:uc-prediction}, the correction improves the 
initial XL-DC structure of 2.9\,mm RMS to a corrected value of
2.1\,mm.

\begin{figure}[t]
  \centering
  \includegraphics[width=0.7\textwidth]{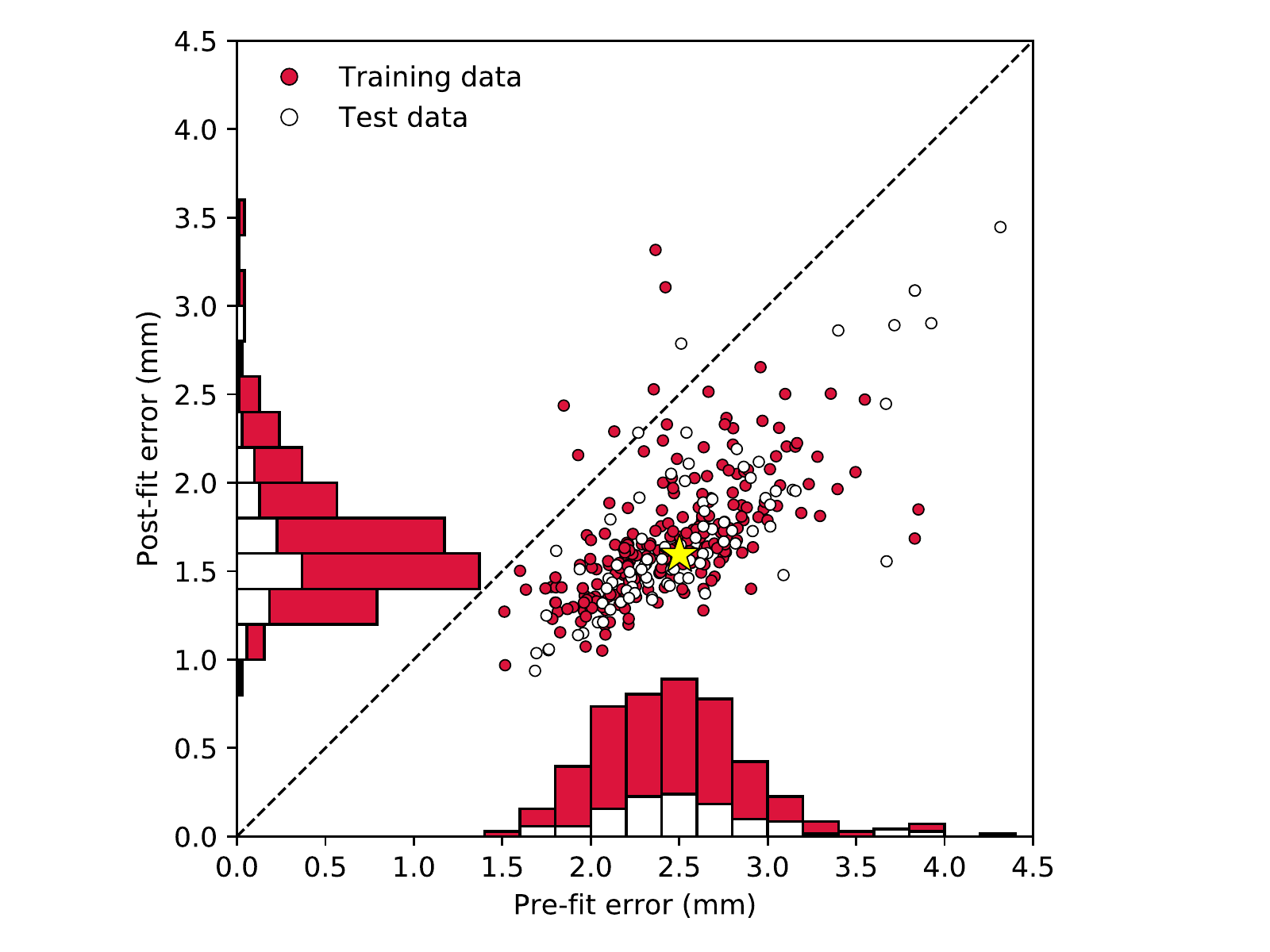}
  \caption{\label{fig:uc-performance} Comparison between pre-fit XL-DC
clock structure and post-fit structure---expressed in millimeters as
described in the text, having subtracted intrinsic variance from the Cs
clock and universal counter.
Each dot represents an
individual 24-hour period. The diagonal dashed line shows where the pre-fit
and the post-fit measures are equal.
Points below the diagonal line
indicate improvement.  The star indicates that a typical improvement is
from 2.5\,mm to 1.6\,mm.}
\end{figure}

The clock-correction algorithm was run on 235 days in a training set and 87
days in a validation (test) set.
As shown by Figure~\ref{fig:uc-performance}, there is no significant
difference between the performances on training data or test data.
Typically, the algorithm reduces the variance from 58 to 33 (in $10^{-24}$
units). Subtracting intrinsic variance, typically this means that real
structure atop intrinsic variance starts at $\sim 42$ and is reduced to
$\sim 17$---mapping to an improvement in standard deviation from $6.5\times
10^{-12}$ to $4\times 10^{-12}$. Phrased in terms of one-way Earth-Moon range,
this reduces the clock-generated error from 2.5\,mm to 1.6\,mm.

\subsection{Masking unreliable periods \label{sec:masking}}

Our goal is to apply the prediction/correction to archival APOLLO normal
point data.
The results of this are addressed in Section~\ref{sec:application}, but
first we must grapple with issues arising from periods in which the GPS
steering data are either absent or of poor quality.

The GPS clock suffers outages at times due to local interference associated
with military testing of GPS unreliability measures.  In many cases, the
clock steering process gets stuck and must be restarted---sometimes after
days of inattention.  Because the clock is in an un-disciplined state
during these periods, we throw out any APOLLO normal point data coincident
with these episodes or in the ensuing recovery periods.  Similarly, lacking
requisite data, we do not attempt to form a prediction during these
periods, or in any period during which the GPS clock's self-reported
frequency offset is larger than about $10^{-10}$.  These episodes
constitute $\sim 6$\% of APOLLO data from 2007--2016.  Finally, in $\sim
2$\% of cases, we do not attempt corrections (but keep the normal point)
when sampling of the GPS clock data during a normal point run was
intermittent so that fewer than half of the expected clock records were
available.

Additionally, we find that the prediction becomes less reliable/effective
during periods when the low-pass-filtered frequency offset exceeds
$10^{-11}$ (using $\sim 0.25$~h time constant).  This is distinct from the
$10^{-10}$ limit in the previous paragraph, above which we do not even
attempt a prediction.  These $>10^{-11}$ filtered excursions are found
roughly 8\% of the time, when padding such instances by a $\pm0.25$~h
buffer.  We keep APOLLO normal points during these episodes, but refrain
from applying a correction---instead inflating the APOLLO normal point
uncertainty accordingly (details below).

To summarize and establish terminology, we \emph{exclude} APOLLO normal
points when the clock is stuck due to interference, is recovering from a
stuck state, or reports frequency offsets higher than $10^{-10}$.  We
\emph{correct} APOLLO normal points when we have enough clock-reported data
during the normal point period and when the filtered frequency offset is
below $10^{-11}$.  We keep \emph{uncorrected} APOLLO normal points and
inflate the uncertainties accordingly for
cases in which either we lack sufficient clock data or we see a
filtered-offset \emph{excursion} $>10^{-11}$.

We face choices in establishing parameters to define the unreliable
periods---specifically: the low-pass filtering time constant (more
aggressive filtering reduces amplitude of excursions); the level of the
filtered excursion; and the buffer pad size.  The first and last in this
list are set according to natural timescales seen in the frequency offset
time series, and the excursion level was selected at $10^{-11}$ in a
balance between performance (which is better the more restrictive the
excursion level) and fraction of normal points impacted.

\subsection{Application to APOLLO data \label{sec:application}}

Considering 2343 APOLLO normal points from 2007 May 24 through 2016
February 12 (prior to the Cs clock), we identify which ones are in
quiet periods when the prediction should be well behaved and apply a
correction.  We also identify problematic normal points.  These include: 142
instances coincident with a stuck or recovering GPS clock, which are
excluded from the data set; 3 for which the frequency offset is
$>10^{-10}$, also excluded from the data set; 44 for which clock statistics
are too sparse; and 193 near an ``offset excursion.'' We keep the last
two categories in the normal point set, but apply an inflated normal
point uncertainty rather than a correction.

\begin{figure}[t]
  \centering
  \includegraphics[width=0.7\textwidth]{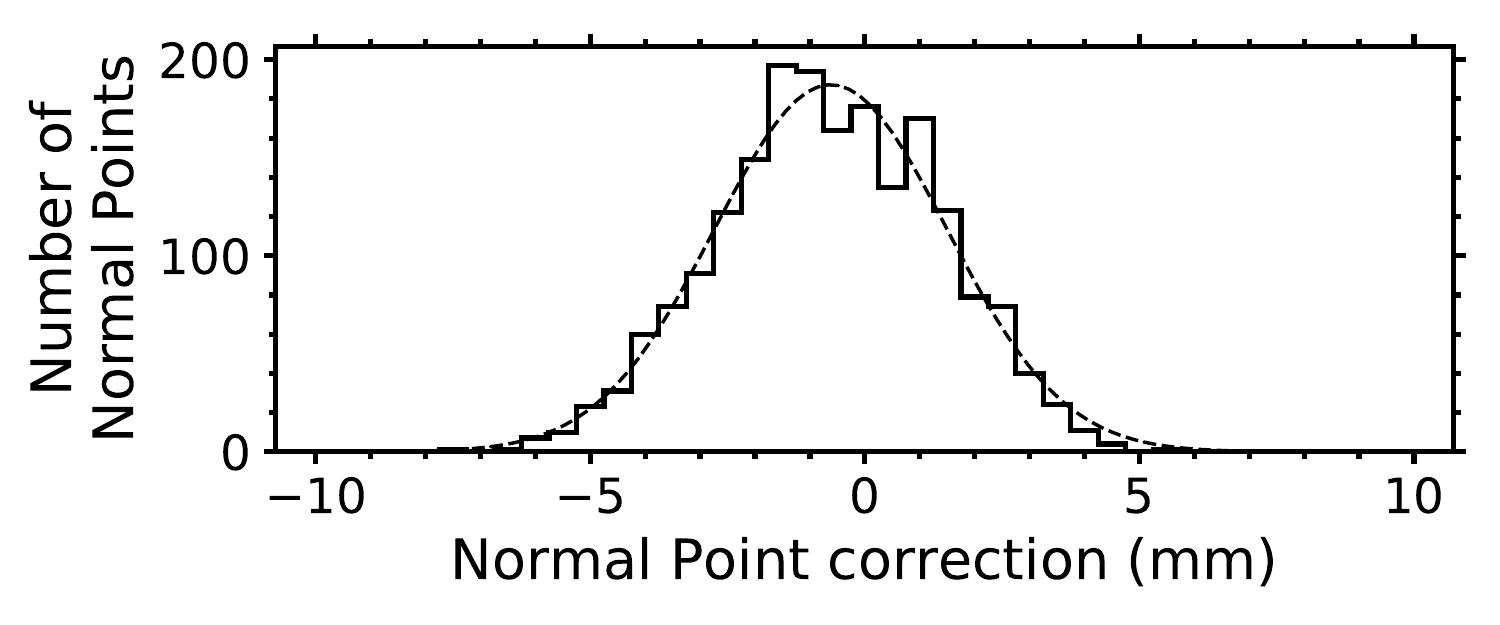}
  \caption{\label{fig:np-corr} Corrections to historical APOLLO normal points based on
the correction algorithm applied to GPS clock data,
in 0.5\,mm bins.  We find an offset of $-0.65$\,mm and standard deviation of 2.0\,mm
(2.1\,mm according to Gaussian fit). }
\end{figure}

The histogram of applied corrections (numbering 1961) appears in
Figure~\ref{fig:np-corr}.  The standard deviation of corrections is $\sim
2$~mm---somewhat lower than the $\sim 3$~mm in Figure~\ref{fig:clk-cmp} or
the pre-fit error of about $\sim 2.5$~mm in
Figure~\ref{fig:uc-performance}, because corrections are not applied to
periods of greater excursion, as indicated above.

Our final task is to arrive at a prescription for assigning systematic
uncertainty contributions in various scenarios.  These uncertainties will
be added to intrinsic APOLLO normal point uncertainties in quadrature to
account for imperfect (or missing) information on clock corrections.  To
assess the degree to which the prediction/correction misses the mark, we
look at the recent year-long period for which we have UC measurements of
the GPS frequency relative to the Cs clock, and compare the prediction to
``truth'' (from the UC) in simulated normal point intervals tiling the
entire year.  As before, we can split the group into points that are
uncorrected due to offset excursions and those for which we can trust the
correction.  For the corrected points, we find that smaller corrections
tend to be more accurate.  Indeed, we can say that the standard deviation
of the prediction error (in mm; relative to the UC ``truth'') behaves
approximately like $1.3+0.04\,x^2$, where $x$ is the prediction/correction
magnitude, in millimeters.  Note that this is not far removed from the
``star" median in Figure~\ref{fig:uc-performance}, and also supports the
trend seen in this figure that smaller pre-fit errors produce smaller
post-fit errors.  At the outer edge of the distribution of corrections seen
in Figure~\ref{fig:np-corr}---namely 5--6~mm---the uncertainty roughly
doubles from its low-end value to $\sim 2.5$~mm.  Meanwhile, we find that
the points that are not corrected due to excessive offset-excursion
($>10^{-11}$ filtered frequency offset) have a roughly constant $\sim
2.7$~mm scatter independent of the correction magnitude that would have
been applied.  Thus the uncertainties for the corrected and uncorrected
points smoothly connect, so that we can apply a sliding-scale error
contribution in quadrature to corrected points, and a ``high-end" error
contribution around 2.7~mm for uncorrected points.

\begin{figure}[t]
  \centering
  \includegraphics[width=0.7\textwidth]{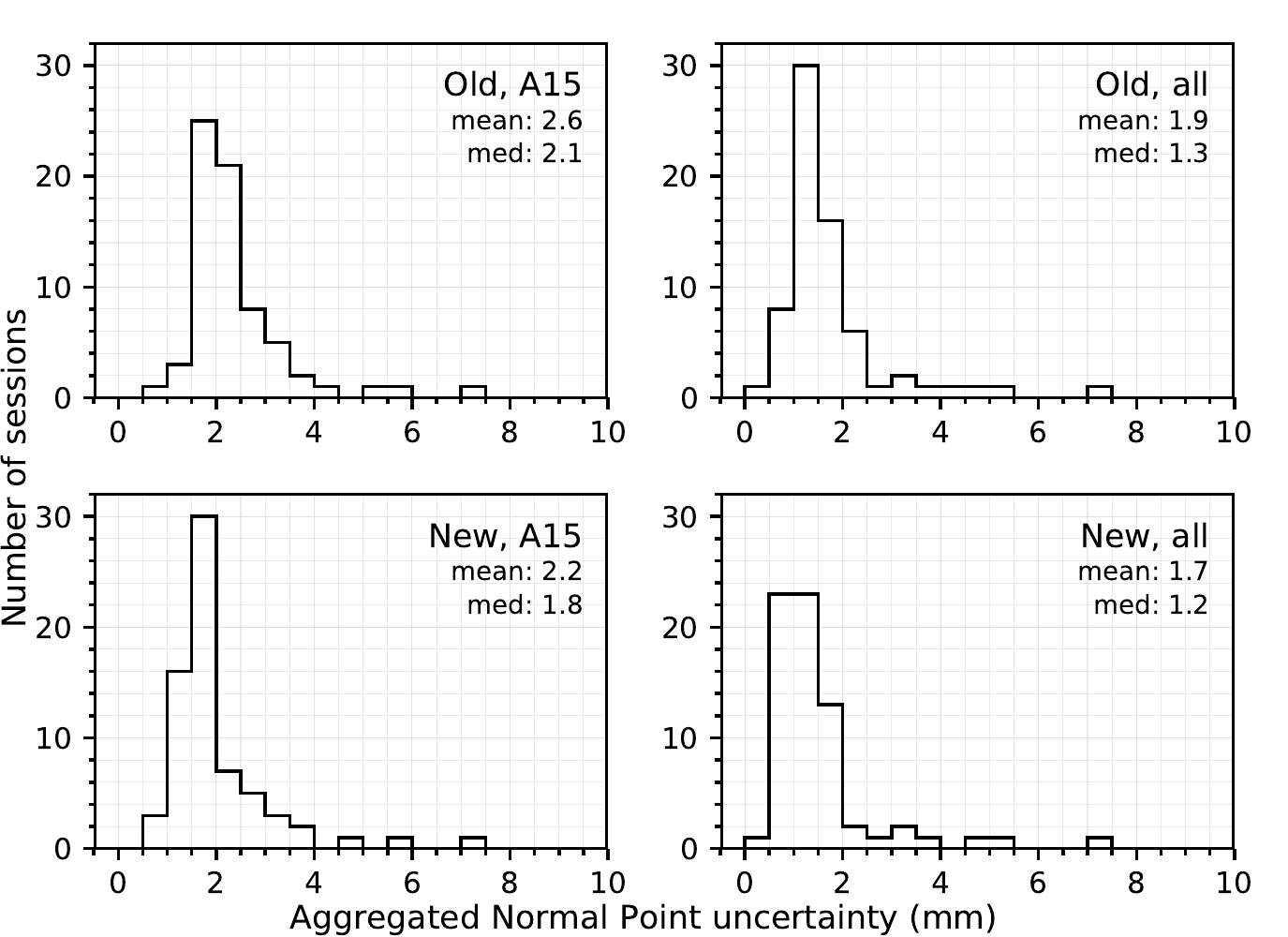}

  \caption{\label{fig:np-result} APOLLO estimated uncertainties for normal
points, aggregated within a session, spanning 69 sessions in recent years
(binned at 0.5~mm).
The top row shows the performance without clock corrections, instead
applying 2.5~mm to each normal point uncertainty in quadrature.  The bottom
row is after correction, reducing uncertainties by the scheme described in
the text.  Results to a single (Apollo~15) reflector are at left, while all
reflectors are aggregated into a representative range error for the entire
session. 41\% of normal point measurements during this period were to the
Apollo~15 reflector (easiest to acquire, especially in poor conditions, and
used to bracket sessions for enhanced sensitivity to Earth orientation). }
\end{figure}

The benefit of the clock error correction on the APOLLO data set is shown in 
Figure~\ref{fig:np-result}.
Because
instrumentation changes since 2006 have resulted in varied performance, it is
practical to isolate contiguous periods of uniform characteristics.  For
this reason we present the latest period of 2013 September 30 to 2016
February 12: after a detector electronics upgrade and before the Cs clock
installation.  In this period, APOLLO had 69 successful sessions (nights) of
operation (good weather, no clock maladies, etc.), producing 562 normal points.  It is
routine for APOLLO to gather multiple measurements to each of the five reflectors
within the course of a session (usually a bit less than one hour in
duration).  The choice to cycle around the reflectors multiple times helps
to evaluate systematics, but since nothing of scientific interest happens
on such short timescales (\ie, relating to gravitational physics), 
we aggregate multiple measurements of a single reflector within one session into a single
representative range uncertainty.  Likewise, we can aggregate
all reflector measurements in a session to produce a single characteristic
measurement uncertainty for the Earth--Moon range at that epoch.  We find that the clock error correction improves the normal point uncertainty distributions and statistics, as seen in
Figure~\ref{fig:np-result}.  This is especially evident at the single-reflector
level, where the number of sessions producing range uncertainties over 2~mm
on Apollo~15 is cut in half from 58\% to 29\%.

\section{Conclusions}

The GPS-disciplined clock historically used as the time and frequency
standard for APOLLO is found to contribute systematic errors in APOLLO data
roughly twice as large as previously estimated---at the level of $\sim
3$~mm in one-way range.  We now have various methods available to correct
APOLLO data in most cases.  In this regard, we can define four periods: an
initial period from 2006 April to 2007 May during which GPS clock statistics
(while collected for most times) were not collected during active periods
of lunar ranging.  For this period, we can offer no correction and instead
inflate the estimated uncertainty by adding $\sim 3.0$~mm in quadrature
with the statistical precision of the normal point.  Next, in the period
from 2007 May to 2016 February, we can correct the vast majority (88\%) of
APOLLO normal points, achieving uncertainties typically around
1.6~mm---again inflating errors by about 2.7~mm when correction is not
possible or ill-advised.  For the period from 2016 February through the end
of 2016, we are able to use independent measurements of the GPS clock
frequency referenced to a Cs standard to apply corrections.
Starting in 2017, APOLLO measurements are based directly on
the Cs frequency standard, obviating the need for corrections to the
GPS clock.  In these final cases, the uncertainty contribution
from the Cs clock derives from the $0.6\times 10^{-12}$ Allan deviation figure at
the $\sim 10^2$~s averaging time for a normal point, yielding a negligible
$\sim 0.3$~mm influence.

We have therefore realized a scheme to make past APOLLO normal points more
accurate, removing much of the deleterious influence from the
GPS-disciplined clock.  Additionally, we can more fairly represent
uncertainty contributions from the clock in various regimes in APOLLO's
past.  Ideally, this will translate to sharper model residuals and an
enhanced ability to effect model improvements.

\subsection*{Acknowledgments}

We thank Russet McMillan for continued excellent service to the APOLLO
project in acquiring data, and the rest of the APOLLO collaboration for
supporting roles.  We thank Ed Leon for performing periodic measurements of
clock phase at the observatory.  This work is based on access to and
observations with the Apache Point Observatory 3.5-meter telescope, which
is owned and operated by the Astrophysical Research Consortium.  This work
was jointly funded by the National Science Foundation (PHY-1404491) and the
National Aeronautics and Space Administration (NNX-15AC51G).

\section*{References}
\bibliographystyle{iopart-num}
\bibliography{apollo}

\end{document}